\documentclass[11pt]{article}
\usepackage{moriond,epsfig,xspace,cite,relsize}

\def\D     {\ensuremath{D}\xspace}

\def\Dbar  {\kern 0.18em\overline{\kern -0.18em D}{}\xspace}
\def\Dzb   {\ensuremath{\Dbar^0}\xspace}

\def\Dstarstar {\ensuremath{D^{**}}\xspace}
\def\Bz    {\ensuremath{B^0}\xspace}
\def\B     {\ensuremath{B}\xspace}
\def\Bbar  {\kern 0.18em\overline{\kern -0.18em B}{}\xspace}

\def\Bzb   {\ensuremath{\Bbar^0}\xspace}

\def\Bs    {\ensuremath{B_s}\xspace}
\def\Bsb   {\ensuremath{\Bbar_s}\xspace}

\def\BzBzb {\ensuremath{B^0 {\kern -0.16em \Bzb}}\xspace}
\def\X     {\ensuremath{X(3872)}\xspace}

\def\babar{\mbox{\slshape B\kern-0.1em{\smaller A}\kern-0.1em
    B\kern-0.1em{\smaller A\kern-0.2em R}}}

\begin{document}
\begin{flushright}
FERMILAB-Conf-04/096-E
\end{flushright}
\vspace*{3.7cm}
\title{\boldmath $B$ Physics at D\O}

\author{ Jan Stark\\ on behalf of the D\O~Collaboration\\[5pt] }

\address{Laboratoire de Physique Subatomique et de Cosmologie, Grenoble, France}

\maketitle\abstracts{
The Fermilab Tevatron ($p\overline{p}$), operating at $\sqrt{s}=1.96$~TeV, is a rich source of \B~hadrons. The large acceptance in terms of rapidity and transverse momentum of the charged particle tracking system and the muon system make the upgraded Run~II D\O~detector an excellent tool for \B~physics. In this article, we report on selected physics results based on the first 250~pb$^{-1}$ of Run~II data. This includes results on the \X~state, semileptonic \B~decays, \B~hadron lifetimes, flavour oscillations, and the rare decay $B_s \rightarrow \mu^+ \mu^-$.
}

\section{Introduction}

The D\O~Collaboration has implemented a major upgrade of their detector for Run~II of the Fermilab Tevatron. Certain components of this upgrade have significantly improved D\O's \B~physics potential compared to Run~I. The upgrade is described in more detail in Ref.~\cite{bib:RunII}. One of the aspects most relevant for \B~physics is the new charged particle tracking system located in a 2~T~field produced by a superconducting solenoidal magnet. The tracking system consists of a scintillating fibre tracker and an inner tracker based on silicon detectors. It provides coverage over a wide range in pseudo-rapidity ($|\eta|<2$, with potential for extension up to $|\eta|<3$) and allows the efficient reconstruction of tracks, including down to low transverse momentum of $p_{\rm T} =180$~MeV/$c$. Due to the limited radius of the fibre tracker (imposed by the geometry of calorimeter, reused from Run~I), the track $p_{\rm T}$-resolution is not as good as, {\it e.g.}, in CDF, which translates into less good resolutions on the mass of fully reconstructed \B~hadrons. Typical mass resolutions for decay modes\footnote{Throughout this article, references to a hadron or to a decay reaction also imply their charge conjugate.} like $B^+ \rightarrow J/\psi(\rightarrow \mu^+ \mu^-) K^+$ are 40~MeV/$c^2$, which still allows a clean identification of \B~hadrons. Track impact parameters~$d_0$, in the plane transverse to the beam axis, with respect to the primary vertex are measured with high precision ($\sigma(d_0) = 50$~$\mu$m for tracks with $p_{\rm T} \simeq 1$~GeV/$c$, improving asymptotically to $\sigma(d_0) = 15$~$\mu$m for tracks with $p_{\rm T} \ge 10$~GeV/$c$), which allows precise measurements of the time-dependence of \B~decays (lifetimes, flavour oscillations, CP~violation, etc). Another key ingredient of the upgraded D\O~detector is the muon detector and its associated trigger system. The muon detector, located outside the calorimeter, relies on three layers of scintillation counters and drift chambers, one of which is placed in front of toroidal magnets, the other two behind the toroids. Muons with momentum greater than $\sim 1.4$~GeV/$c$ ($\sim 3.5$~GeV/$c$) penetrate the calorimeter (toroid magnet) at $\eta = 0$. The muon system provides wide coverage over $|\eta|<2.0$, including at the trigger level. The data used for the measurements reported here have been recorded with single and di-muon triggers. The single muon triggers are prescaled away at the highest instantaneous luminosities.\\
Most of the results reported in this article are based on 250~pb$^{-1}$ of data (before prescales) collected during the period from April~2002 to January~2004. Some of the results are based on earlier subsets of this data sample.

\section{\boldmath Production and decay characteristics of the $X(3872)$}

\begin{figure}
\begin{center}
\epsfig{file=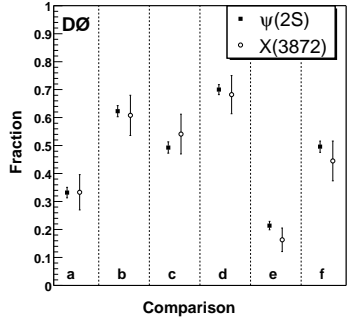,width=0.3\textwidth}
\caption{Comparison of event-yield fractions for \X and $\psi(2S)$ in the regions: (a)~$p_{\rm T}(J/\psi \pi^+ \pi^-) > 15 \; {\rm GeV}/c$; (b)~$|y(J/\psi \pi^+ \pi^-)| < 1$; (c)~$\cos(\theta_{\pi})<0.4$; (d)~effective proper decay length, $dl < 0.01$~cm; (e)~isolation~$=$~1; (f)~$\cos(\theta_{\mu})<0.4$~.}
\label{fig:X}
\end{center}
\end{figure}

The Belle Collaboration has recently discovered a new particle, the \X, with a mass of $3872.0 \pm 0.6 \, ({\rm stat}) \pm 0.5 \, ({\rm syst})$~MeV/$c^2$, produced in $B^{\pm} \rightarrow \X K^{\pm}, \; \X \rightarrow J/\psi \pi^+ \pi^-$ decays~\cite{bib:XBelle}. The existence of the \X~state has been confirmed, also using the $J/\psi \pi^+ \pi^-$ mode, in $p\overline{p}$~collisions by CDF~\cite{bib:XCDF}. At this time, it is still unclear whether this particle is a $c\overline{c}$~state or a more complex object; see, {\it e.g.}, Refs.~\cite{bib:XEichten,bib:XQuigg,bib:XToern}.\\
We perform a more detailed study~\cite{bib:XD0} of the production and decay characteristics of the \X in $p\overline{p}$~collisions. The $c\overline{c}$~state $\psi(2S)$ has a mass $m(\psi(2S)) = 3685.96 \pm 0.09$~MeV/$c^2$~\cite{bib:PDG} close to the \X, and has the same decay mode~$J/\psi \pi^+ \pi^-$. It therefore provides a good benchmark for comparison with the~\X. Using 230~pb$^{-1}$ of data, we reconstruct $522 \pm 100$ $\X \rightarrow J/\psi \pi^+ \pi^-$ decays. We split our samples of $\psi(2S)$ and \X decays into two subsamples according to transverse momentum $p_{\rm T}(J/\psi \pi^+ \pi^-)$, and compare the event-yield fractions $f_{\psi(2S)}$ and $f_{\X}$ with
$$
f_{\rm P} = \frac{N(P) {\rm ~at~} p_{\rm T}(J/\psi \pi^+ \pi^-) > 15 \; {\rm GeV}/c^2}{N(P) {\rm ~at~} p_{\rm T}(J/\psi \pi^+ \pi^-) \le 15 \; {\rm GeV}/c^2} \; ,
$$
where $P=\psi(2S)$ or~\X. The result is shown in Fig.~\ref{fig:X}. We repeat this study for splittings according to the rapidity~$y$ of the $J/\psi \pi^+ \pi^-$ system, the helicity of the $\pi^+ \pi^-$ ($\mu^+ \mu^-$) system, the effective proper decay length and isolation~(Fig.~\ref{fig:X}). The helicity of the $\pi^+ \pi^-$ ($\mu^+ \mu^-$) system is inferred by boosting one of the pions (muons) and the $J/\psi \pi^+ \pi^-$ system into the di-pion (di-muon) rest frame and measuring the angle~$\theta_{\pi}$~($\theta_{\mu}$) between them. The effective proper decay length~$dl$ is defined as the distance in the transverse plane from the primary vertex to the decay vertex of the~$J/\psi$, scaled by the mass of the $J/\psi \pi^+ \pi^-$ system divided by $p_{\rm T}(J/\psi \pi^+ \pi^-)$. The isolation is defined as the ratio of the $J/\psi \pi^+ \pi^-$ momentum to the sum of the momentum of the $J/\psi \pi^+ \pi^-$ and the momenta of all other reconstructed charged particles within a cone of radius $\Delta R = 0.5$ in $\eta/\phi$~space about the direction of the $J/\psi \pi^+ \pi^-$ momentum. In summary, no significant differences between the \X and the $c\overline{c}$~state $\psi(2S)$ are found when the data are separated according to any of these production and decay variables.

\section{\boldmath Precision measurement of $\tau(B^+)/\tau(B^0)$}

\begin{figure}
\hfill
\begin{minipage}[t]{0.45\textwidth}
\begin{center}
\epsfig{file=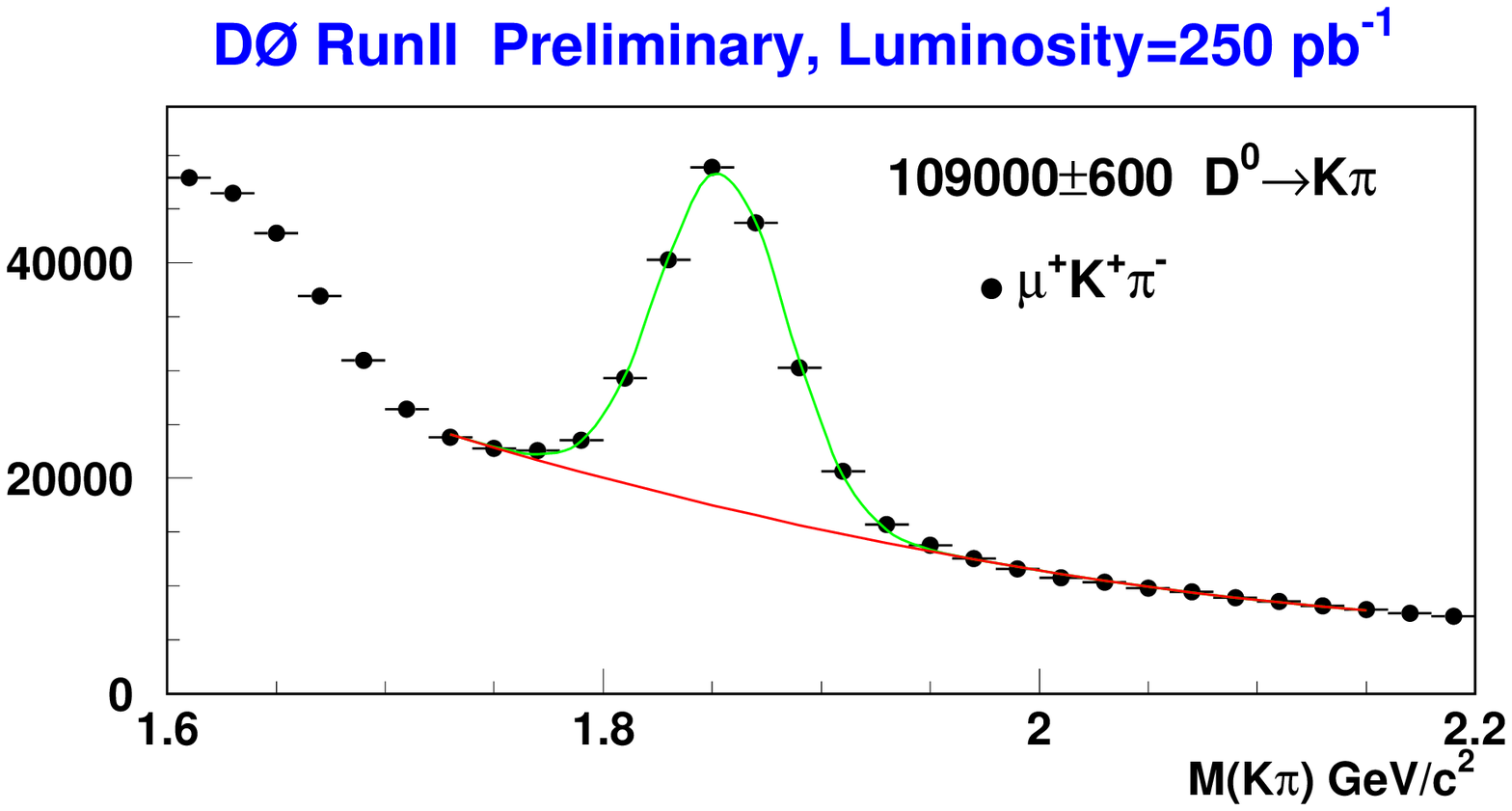,scale=0.42}
\caption{Invariant mass $m(K^+ \pi^-)$ of the $K^+ \pi^-$~system in $K^+ \pi^- \mu^+$~events. The solid lines represent the result of a fit to the sum of a gaussian distribution to model the $\Dzb \rightarrow K^+ \pi^-$ peak and a polynomial background model. Data points in the low mass peak, corresponding to partially reconstructed $\Dzb \rightarrow K^+ \pi^- X$ decays, are assigned a zero weight in this fit.}
\label{fig:tauD0}
\end{center}
\end{minipage}
\hfill
\begin{minipage}[t]{0.45\textwidth}
\begin{center}
\epsfig{file=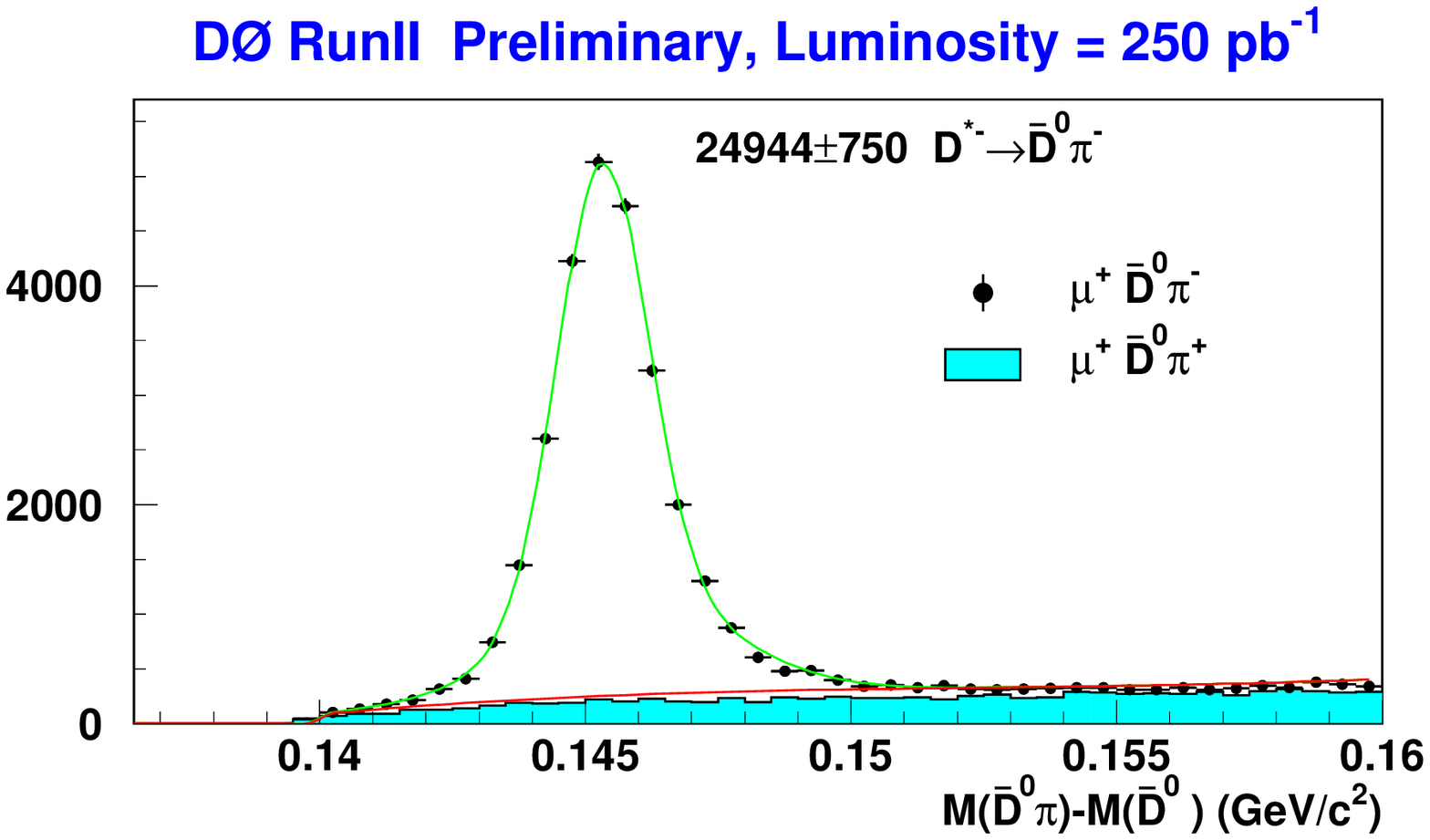,scale=0.40}
\caption{Mass difference $m(\Dzb \pi^-) - m(\Dzb)$ for $\Dzb \pi^- \mu^+$ events with $1.80 < m(\Dzb) < 1.90$~GeV/$c^2$.}
\label{fig:tauDst}
\end{center}
\end{minipage}
\hfill
\end{figure}

The spectator quark model predicts that the two charge states with one heavy quark~$Q$ ($Q\overline{u}$ and $Q\overline{d}$) have the same lifetime. Deviations from this simple picture have been evaluated from first principles of QCD using Operator Product Expansion (an overview is presented, {\it e.g.}, in Ref.~\cite{bib:BigiOPE} and references therein) and are expected to be proportional to~$1/[m(Q)]^3$. Using inputs from quenched lattice QCD, recent NLO calculations~\cite{bib:NiersteTau} yield $\tau(B^+)/\tau(B^0) = 1.053 \pm 0.016 \pm 0.017$, but unquenching effects could be sizeable.\\
We measure $\tau(B^+)/\tau(B^0)$ in 250~pb$^{-1}$ of data using large samples of semileptonic $B^+$ and \Bz~decays collected using single muon triggers. Semileptonic \B~decays are reconstructed in the final state~$\Dzb(\rightarrow K^+ \pi^-)\mu^+ X$, where $X$ denotes any additional decay products. Figure~\ref{fig:tauD0} shows the $m(K^+ \pi^-)$~spectrum for $K^+ \pi^- \mu^+$~events after $p_{\rm T}$~and tracking hit requirements on the tracks and cuts on the significance of the impact parameter of the $K^+$ and $\pi^-$ tracks, the significances of the distances in the transverse plane between the primary vertex and the \Dzb and \B vertices, the angle in the transverse plane between the $\Dzb \mu^+$ momentum and the direction of the $\Dzb \mu^+$ vertex w.r.t. the primary vertex, as well as the requirement $2.3 < m(\Dzb \mu^+) < 5.2$~GeV/$c^2$. The signal in the \Dzb~peak contains $109k$~events. The next step is to identify the subsample of events in which the \Dzb originates from a $D^{*-} \rightarrow \Dzb \pi^-$~decay. Due to the small $D^{*-}$/\Dzb mass difference, the transverse momentum of the pion tends to be low, and for small transverse momentum of the $D^{*-}$, the slow pion reconstruction efficiency is limited by tracking acceptance. For each $\Dzb \mu^+$ candidate, an additional pion with charge opposite to the charge of the muon is searched for. The mass difference $\Delta m = m(\Dzb \pi^-) - m(\Dzb)$ for events with such an additional slow pion is shown in Fig.~\ref{fig:tauDst}. The peak corresponds to $D^{*-} \mu^+$ events.\\
The slow pion reconstruction allows the sample to be split into a ``\Dzb~sample'' ($\Dzb(\rightarrow K^+ \pi^-)\mu^+ X$ events in which no slow pion is found) and a ``$D^*$~sample'' ($\Dzb(\rightarrow K^+ \pi^-) \pi^- \mu^+ X$). The \Dzb~sample is enriched in $B^+$~decays, while the $D^*$~sample is enriched in~\Bz. This separation between the \B~species gives us access to the lifetime ratio~$\tau(B^+)/\tau(B^0)$. We further require $p_{\rm T}(\Dzb)> 5$~GeV/$c$. This cut reduces the event yields by $\sim 35$~\%, but it also ensures a good slow pion reconstruction efficiency (88~\%, independent, within uncertainty, of the \B~meson decay length) and therefore a good separation between the \Dzb~and~$D^*$ samples. The relative abundances of different species of \B~hadrons in the \Dzb~and~$D^*$ samples are evaluated from measured branching fractions and their uncertainties, and the reconstruction efficiencies for the different decay chains. Isospin relations are used to infer some unmeasured branching fractions from measured ones. The \Dzb~sample contains 82~\%~$B^+$, 16~\%~\Bz and 2~\%~\Bs, while the $D^*$~sample contains 86~\%~\Bz, 12~\%~$B^+$ and 2~\%~\Bs.\\
As the semileptonic \B~decays in our samples are only partially reconstructed, the proper decay time of a \B~candidate can only be inferred approximately from the measured decay length in the transverse plane~$l_{\rm T}$. For each event, we calculate the {\it visible proper decay length} (VPDL) defined as $x^m = l_{\rm T}\cdot\frac{m(B)}{p_{\rm T}(\Dzb\mu)}$, and we require $\sigma(x^m) < 200$~$\mu$m. The expected VPDL distribution for a given \B~species and a given decay chain is obtained from the expected proper decay time distribution and the distribution of $K=p_{\rm T}(\Dzb\mu)/p_{\rm T}(B)$ (``$K$-factor'', determined from simulated events). The lifetime ratio $\tau(B^+)/\tau(B^0)$ could be extracted from fits to the VPDL distributions of the \Dzb~and~$D^*$ samples. We chose a different technique. We group the events into 8~bins of VPDL, and measure the ratio 
$$
r_i = \frac{N_i(D^{*-}\mu^+)}{N_i(\Dzb\mu^+)}
$$ 
separately in each bin~$i$. The event yields $N_i(D^{*-}\mu^+)$ and $N_i(\Dzb\mu^+)$ are determined from separate fits to the $m(K^+\pi^-)$ mass spectra (similar to the one in Fig.~\ref{fig:tauD0}) of the \Dzb~and~$D^*$ samples in each bin. The $D^*$~sample contains a small contribution of combinatorial background from events with a true~\Dzb, which is counted as signal in the $m(K^+\pi^-)$~fit. This contribution is subtracted using the $m(K^+\pi^-)$ distribution of wrong charge combinations~$\Dzb \pi^+$ in each VPDL bin and a correction factor $C$ to take into account differences in the combinatorial background between right and wrong charge combinations. The correction factor $C = 1.22 \pm 0.04$ is derived from the full sample. The lifetime ratio $\tau(B^+)/\tau(B^0)$ is determined from a fit to the 8~values~$r_i$. An advantage of this technique compared to fits to VPDL distributions is that the background is properly subtracted in all VPDL regions and no parameterisation of the VPDL distribution of background is needed in the fit. This technique also provides an elegant way to take into account the decay length significance cuts described above. Additional inputs to the $r_i$~fit are the relative \B~contents of the \Dzb~and~$D^*$ samples, taking into account the effect of reconstruction efficiencies (in particular the slow pion reconstruction efficiency), the distributions of $K$-factors for different decay chains, the decay length resolution (from simulation) and the lifetime~$\tau(B^+)$ which is fixed to the PDG value $\tau(B^+) = 1.674 \pm 0.018$~ps~\cite{bib:PDG}.\\
Figure~\ref{fig:VPDL} shows the measured values of the~$r_i$, as well as the result of the fit. We obtain the preliminary measurement
$$
\tau(B^+)/\tau(B^0) = 1.093 \pm 0.021 \, ({\rm stat}) \pm 0.022 \, ({\rm syst}) \; .
$$
The systematic uncertainties are dominated by the uncertainty on a possible decay-time dependence of the slow pion reconstruction efficiency, uncertainties in the relative reconstruction efficiencies for different \B~hadron decay chains, the uncertainty in~Br$(B^+ \rightarrow D^{*-} \pi^+ \mu^+ \nu X)$ which is used in the estimate of the sample compositions, the $K$-factors from simulation and the uncertainty in the decay length resolution differences between the \Dzb~and~$D^*$ samples. This preliminary D\O~measurement is one of the most precise measurements of $\tau(B^+)/\tau(B^0)$, and it is consistent with previous measurements~\cite{bib:PDG}. It is also consistent with the most recent preliminary CDF measurement, $\tau(B^+)/\tau(B^0) = 1.080 \pm 0.042$, based on fully reconstructed hadronic $B \rightarrow J/\psi X$ decays in 240~pb$^{-1}$~\cite{bib:Jonas}.

\section{\boldmath \Bz/\Bzb mixing with opposite-side muon tagging}

\begin{figure}
\hfill
\begin{minipage}[t]{0.45\textwidth}
\begin{center}
\epsfig{file=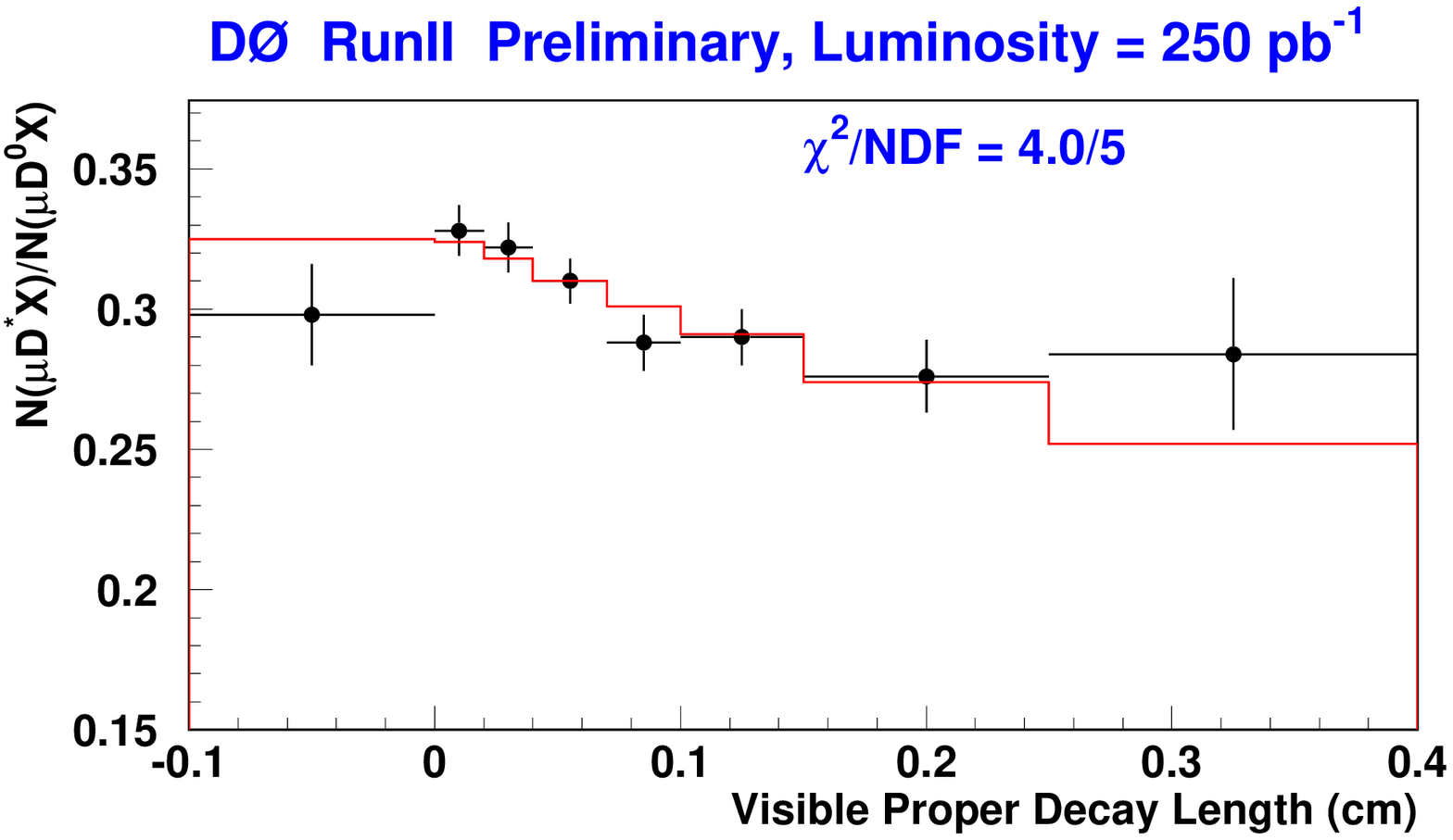,scale=0.37}
\caption{Measured values of $r_i = \frac{N_i(D^{*-}\mu^+)}{N_i(\Dzb\mu^+)}$ in the 8~bins of visible proper decay length (VPDL). The solid histogram represents the result of the fit to the data.}
\label{fig:VPDL}
\end{center}
\end{minipage}
\hfill
\begin{minipage}[t]{0.45\textwidth}
\begin{center}
\epsfig{file=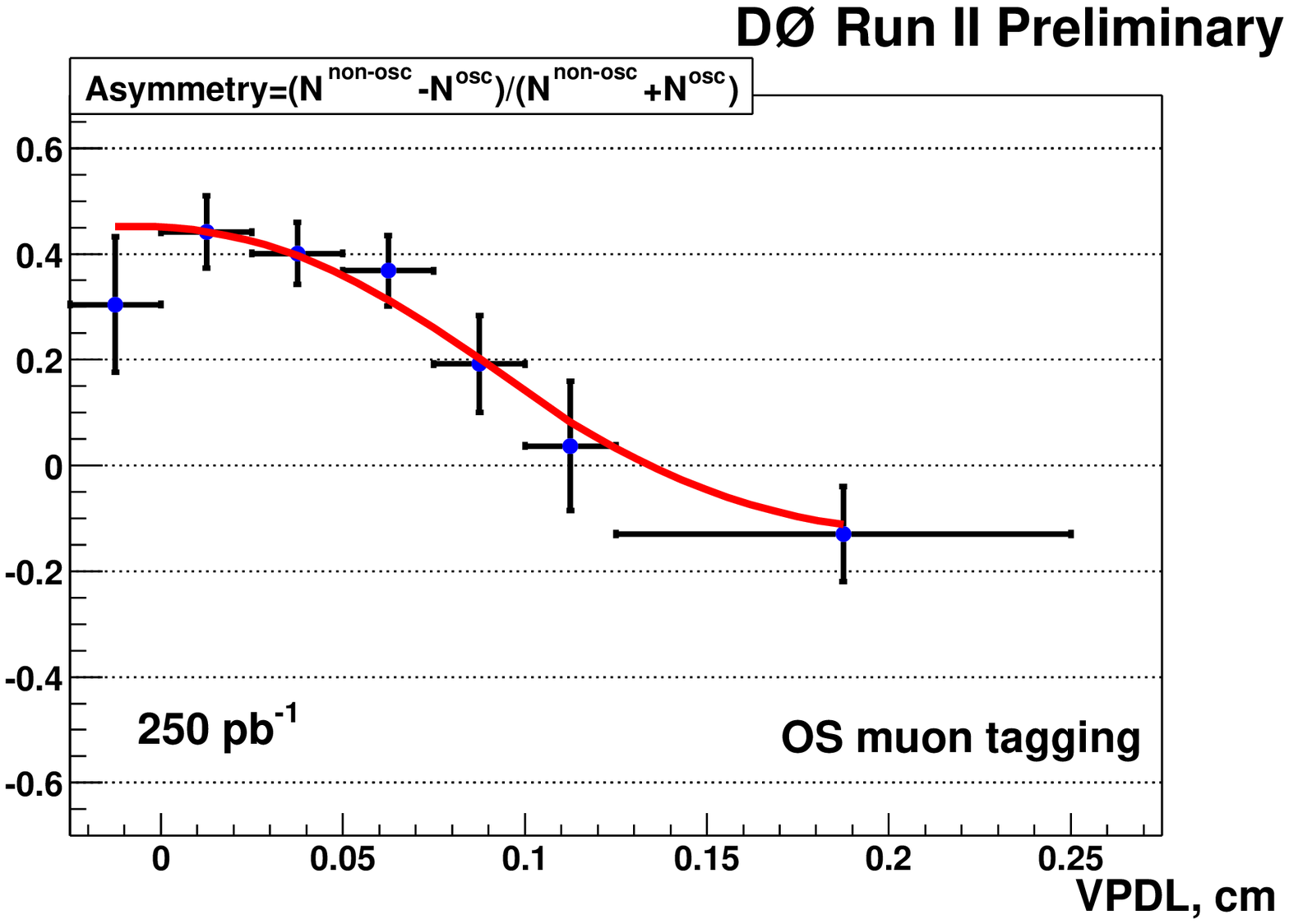,scale=0.37}
\caption{Asymmetry between the number of mixed ($N^{\rm osc}$) and the number of unmixed ($N^{\rm non-osc}$) events as a function of visible proper decay length (VPDL). The solid line represents the result of the fit that is used to measure $\Delta m_d$ and the tagging purity~$\kappa$.}
\label{fig:mixing}
\end{center}
\end{minipage}
\hfill
\end{figure}

We use our sample of semileptonic $\Bz \rightarrow \D^{*-} \mu^+ \nu$ decays to measure the \Bz/\Bzb mixing frequency~$\Delta m_d$. Compared to the lifetime measurement, we need an additional key ingredient: the ability to identify (``tag'') the initial flavour of the \B~meson. Opposite-side muon tagging relies on the charge of muons among the decay products from the ``other~$b$'' from $b\overline{b}$~pair production to infer the initial flavour of the reconstructed~\B. Typically, the $b\overline{b}$~pair leads to two-jet events with two \B~mesons in two back-to-back jets. The final state can also contain a $b$~baryon and/or more than two jets. In addition, if the two $b$-jets originate from the flavour excitation or gluon splitting processes, the angle between the two jets is not necessarily close to 180~degrees, but varies within a rather wide range. We select tagging muons with $p_{\rm T}>2.5$~GeV/$c$ and $\cos(\Delta\phi) < 0.5$, where $\Delta\phi$~is the difference between the $\phi$~directions of the momentum of the $\Dzb \mu^+$~system from the reconstructed~\B and the momentum of the tagging muon. This algorithm provides an estimate of the initial flavour for $(4.76 \pm 0.19)$~\% of all reconstructed \Bz/\Bzb~mesons.\\
The mixing frequency~$\Delta m_d$ has been measured with high precision by many experiments, most recently at the \B-factories. We use the study of \Bz/\Bzb~mixing to benchmark our flavour tagging. A fit to the observed asymmetry between the number of mixed and unmixed events as a function of VPDL allows us a simultaneous measurement of $\Delta m_d$ and the tagging purity $\kappa = \frac{N_{\rm R}}{N_{\rm R}+N_{\rm W}}$, where $N_{\rm R}+N_{\rm W}$ is the number of events for which the tagging algorithm has provided an estimate of the initial flavor, and $N_{\rm R}$ is the number of events in which the estimate was correct. This allows us to measure the performance of our flavour tagging in data before we use it in more difficult studies of, {\it e.g.}, \Bs/\Bsb~mixing and CP~violation. The measurement of~$\Delta m_d$ also constrains more exotic models of $b$~production in hadron collisions (see, {\it e.g.}, Ref.~\cite{bib:Exotic}).\\
Using 250~pb$^{-1}$ of data, we obtain the VPDL-dependent asymmetry between mixed and unmixed events shown in Fig.~\ref{fig:mixing} and the preliminary measurements
$$
\Delta m_d = 0.506 \pm 0.055 \, ({\rm stat}) \pm 0.049 \, ({\rm syst}) \, {\rm ps}^{-1} {\rm ~~~~~and~~~~~} \kappa = 73.0 \pm 2.1 \, ({\rm stat}) \pm 0.8 \, ({\rm syst}) \, \% \; .
$$
The systematic uncertainties on~$\Delta m_d$ are dominated by the uncertainties in the method used to fit the number of signal events in each VPDL bin. The systematic uncertainties on~$\kappa$ are dominated by the uncertainty in the \Bs~contribution to the signal sample, on the proper decay time resolution and in the alignment of the tracking system. The measured value of~$\Delta m_d$ is consistent with the world average~\cite{bib:PDG}.\\
This kind of analysis is currently being repeated for other, complementary, tagging algorithms based on opposite-side electrons, jet charge and same-side tags. These additional taggers will further improve our overall flavour tagging efficiency.

\section{\boldmath Observation of semileptonic \B~decays to narrow \Dstarstar mesons}

Orbitally excited states of the $D$~meson are usually referred to as \Dstarstar~mesons. In the simplest case, \Dstarstar~mesons consist of a charm quark and a light quark in a state with orbital angular momentum~$L=1$. In the limit~$m(c) \gg \Lambda_{\rm QCD}$, the spin of the charm quark decouples from the other angular momenta, and the angular momentum sum $j = S + L$ of the light quark spin~$S$ and the orbital angular momentum~$L$ is conserved. For $L=1$, one therefore expects one doublet of states with~$j = 3/2$ and another doublet with~$j = 1/2$. The two $j = 3/2$ states are denoted $D^0_1$~and~$D^{*0}_2$. In contrast to the $j = 1/2$ states, the $D^0_1$~and~$D^{*0}_2$ decay through a D-wave and are thus expected to be narrow (20~to~30~MeV/$c^2$, while the $j = 1/2$ states decay through an S-wave and are expected to be several hundred~MeV/$c^2$ wide). For a recent theoretical overview of the topic, see, {\it e.g.}, Ref.~\cite{bib:Zoltan} and references therein.\\
The $D^0_1$~and~$D^{*0}_2$ have been observed and studied in several experiments, most recently by Belle~\cite{bib:DststBelle} and \babar~\cite{bib:DststBaBar} in the hadronic decay $B^- \rightarrow D^{**0} \pi^-$. We study $D^0_1$, $D^{*0}_2$ produced in semileptonic \B~decays. Our measurements further constrain \D~meson spectroscopy and improve the understanding of the samples of semileptonic \B~decays we use as a tool for studies of the time-dependence of \B~decays (lifetimes, mixing, etc). Figure~\ref{fig:Dstst} shows the $m(D^* \pi)$ spectrum we observe in candidate $B \rightarrow (D^0_1, D^{*0}_2) \mu^- \nu X \rightarrow D^{*+} \pi^{-} \mu^- \nu X$ decays using 250~pb$^{-1}$ of data. The excess in right-sign combinations can be interpreted as combined effect of the $D^0_1$~and~$D^{*0}_2$. From topological analyses at LEP~\cite{bib:PDG}, we know the inclusive rate
$$
{\rm Br}(B \rightarrow D^{*+} \pi^- \mu^- \nu X) = 0.48 \pm 0.10 \, \% \; .
$$
Our preliminary result constrains the resonant contribution
$$
{\rm Br}(B \rightarrow \{D^0_1, D^{*0}_2\} \mu^- \nu X) 
  \cdot {\rm Br}(\{D^0_1, D^{*0}_2\} \rightarrow D^{*+} \pi^-) 
     = 0.280 \pm 0.021 \, ({\rm stat}) \pm 0.088 \, ({\rm syst}) \, \% \; .
$$
The branching ratio for \B~decays to \Dstarstar has been determined using a normalisation to the decays $B \rightarrow D^{*+} (\rightarrow D^0 \pi^+) \mu^- \nu X$, whose branching ratios are well known~\cite{bib:PDG}. The systematic uncertainties are dominated by the uncertainty on the $p_{\rm T}$~dependence of the soft pion reconstruction efficiency, as well as the effects of possible contributions from wide resonances and the effect of possible interference effects between the $D^0_1$~and~$D^{*0}_2$ on the tails of the \Dstarstar mass distribution, which are not taken into account in this preliminary analysis.

\section{\boldmath Fully reconstructed \B~decays}

\begin{figure}
\begin{center}
\epsfig{file=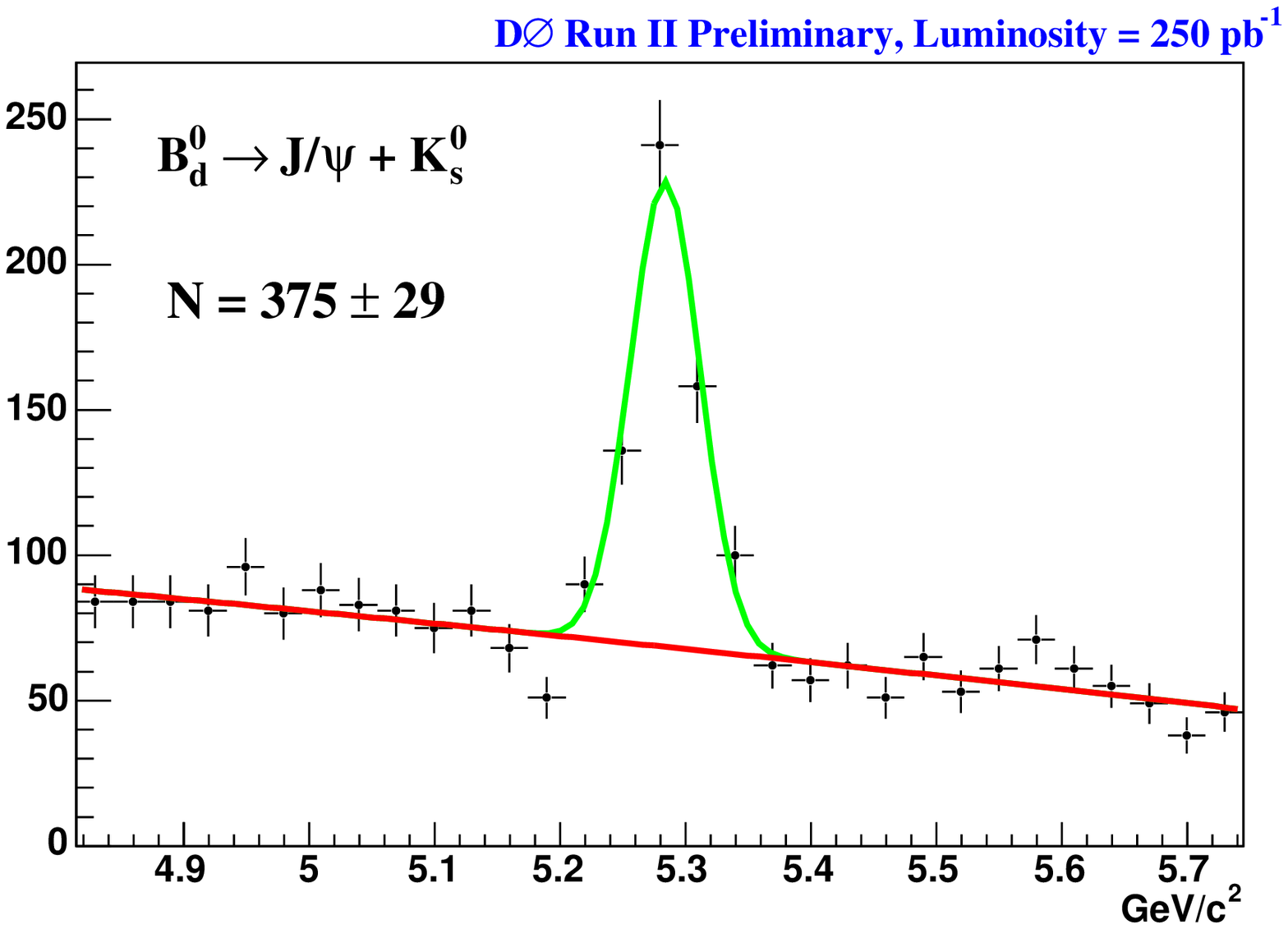,width=0.32\linewidth}
\epsfig{file=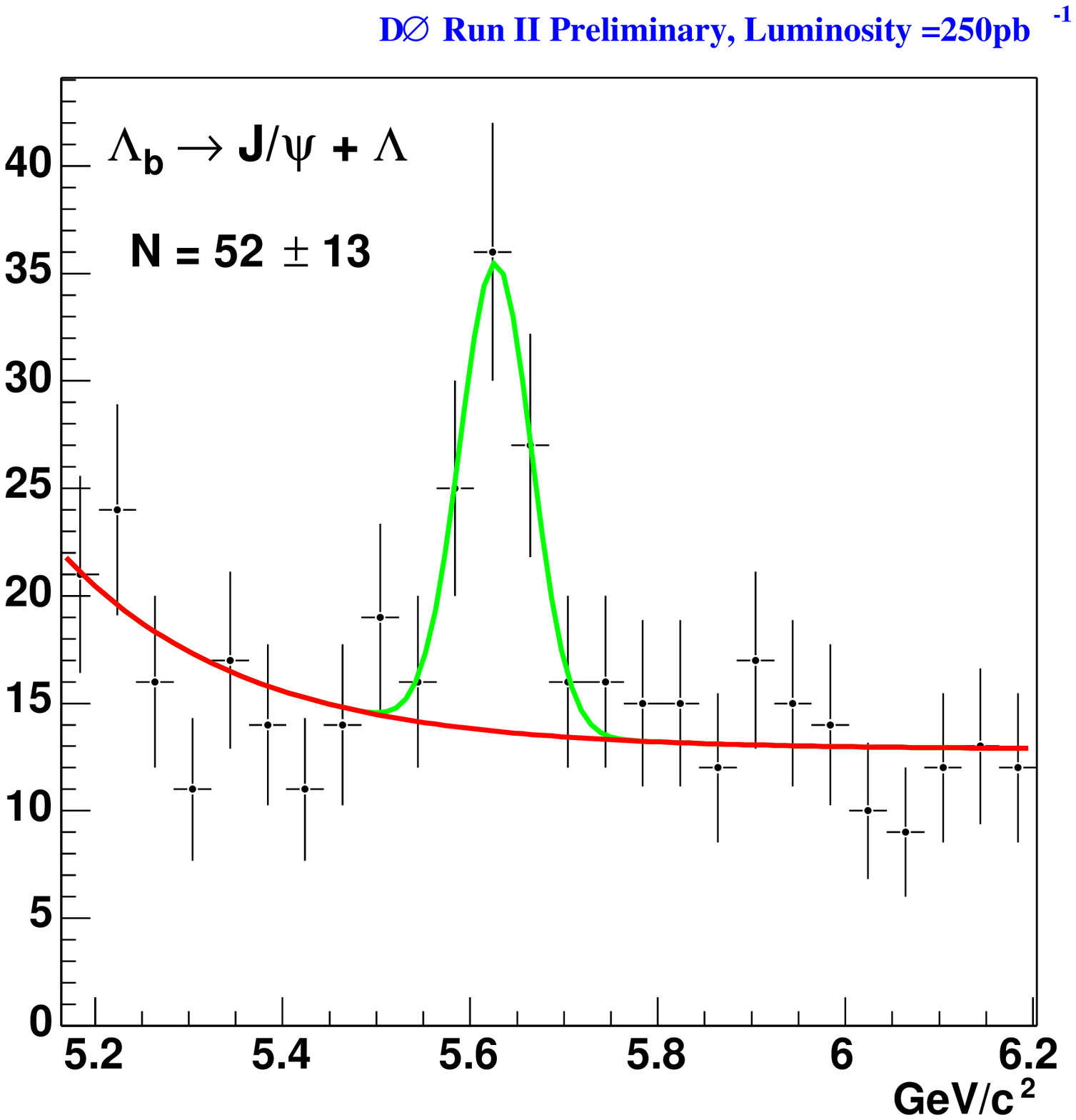,width=0.32\linewidth}
\epsfig{file=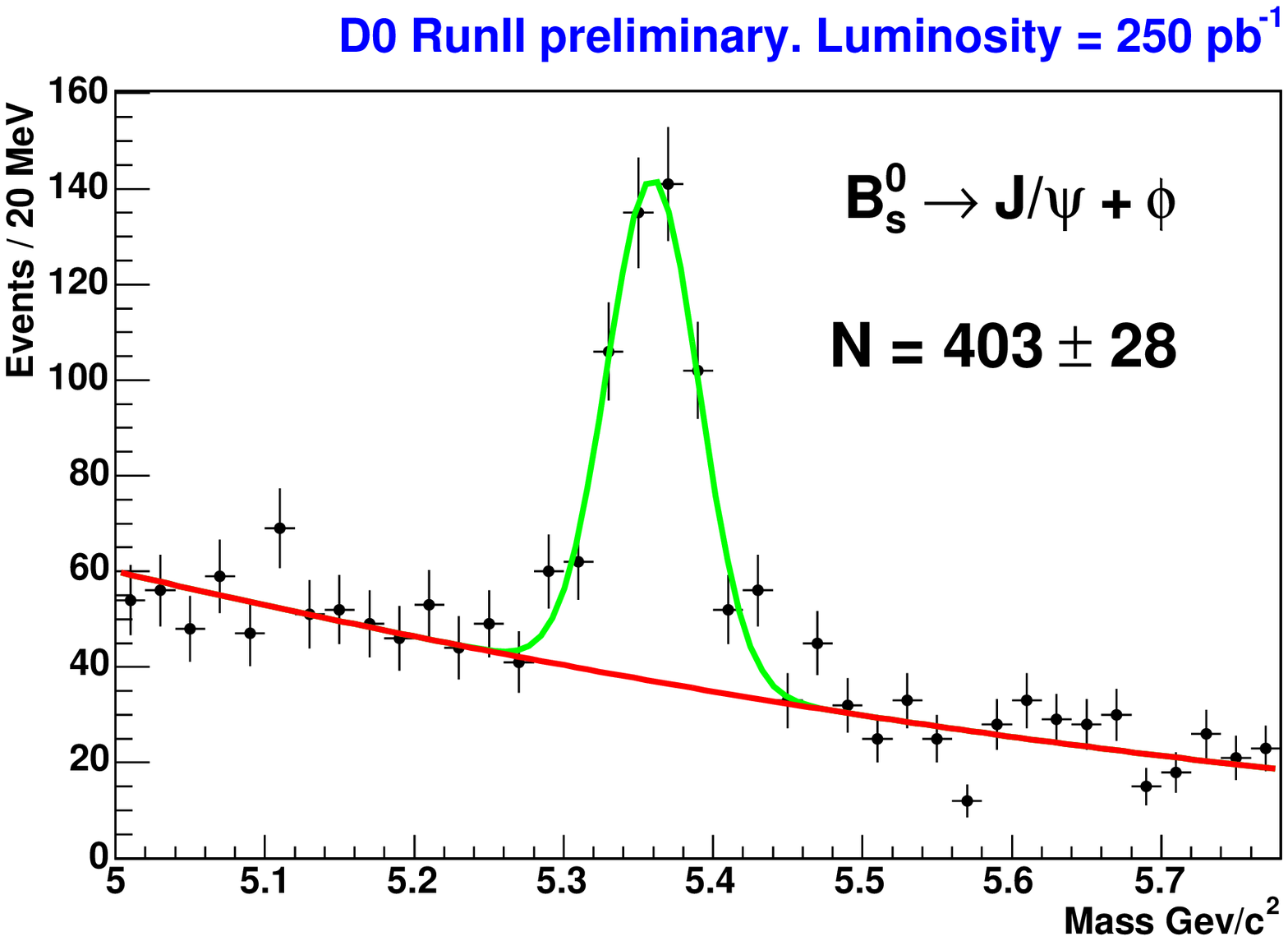,width=0.32\linewidth}
\caption{Fully reconstructed $b$~hadrons in three selected decay modes.}
\label{fig:exclusive}
\end{center}
\end{figure}

D\O's good tracking acceptance and efficiency, including for tracks from, {\it e.g.}, $K^0_{\rm S} \rightarrow \pi^+ \pi^-$ or $\Lambda \rightarrow p^+ \pi^-$ that originate far away from the primary vertex, also allow the exclusive reconstruction of hadronic \B~decays. The example of $\Bz \rightarrow J/\psi(\rightarrow \mu^+ \mu^-) K^0_{\rm S}$ is shown in Fig.~\ref{fig:exclusive}. Using this decay mode in 114~pb$^{-1}$ of data, we obtain the preliminary measurement $\tau(\Bz) = 1.56^{+0.32}_{-0.25} \, ({\rm stat}) \pm 0.13 \, ({\rm syst})$~ps, consistent with the world average~\cite{bib:PDG}. Figure~\ref{fig:exclusive} also shows the $\Lambda_b \rightarrow J/\psi(\rightarrow \mu^+ \mu^-) \Lambda(\rightarrow p^+ \pi^-)$ signal. A lifetime measurement in this channel is currently being finalised. Another signal with rich physics potential, $\Bs \rightarrow J/\psi(\rightarrow \mu^+ \mu^-) \phi (\rightarrow K^+ K^-)$, is also included in Fig.~\ref{fig:exclusive}.

\section{\boldmath Sensitivity analysis of the rare decay $B_s \rightarrow \mu^+ \mu^-$}

\begin{figure}
\hfill
\begin{minipage}[t]{0.5\textwidth}
\begin{center}
\epsfig{file=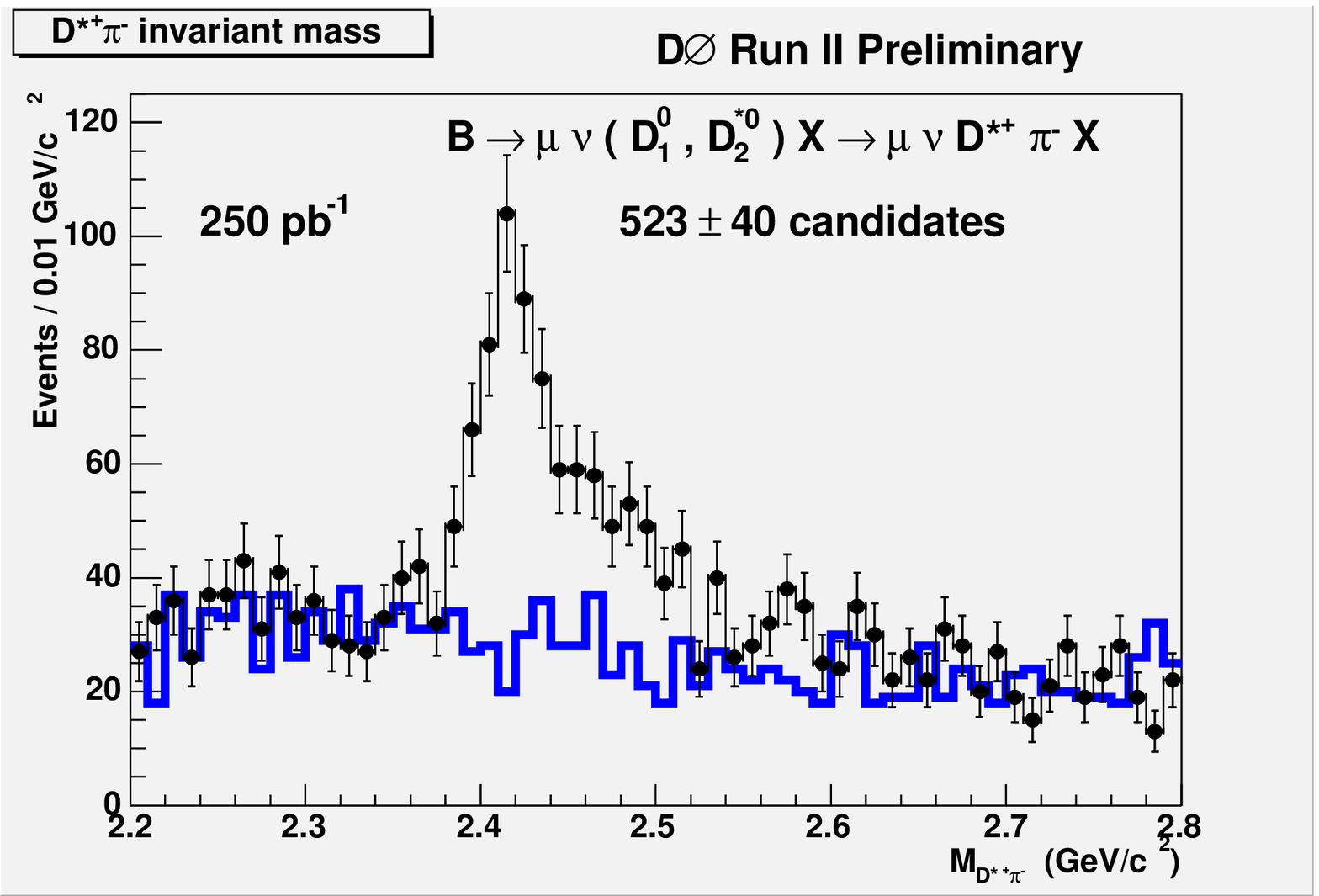,scale=0.40}
\caption{Invariant mass $m(D^* \pi)$ for candidate $B \rightarrow (D^0_1, D^{*0}_2) \mu^- \nu X \rightarrow D^{*+} \pi^{-} \mu^- \nu X$ events. The points correspond to $D^* \pi$ combinations with opposite charges, and the solid histogram corresponds to same charge combinations. The total number of events in the peak is $523 \pm 40$, estimated as difference in the number of events for right charge and wrong charge combinations with $2.35 < m(D^* \pi) < 2.55$~GeV/$c^2$.}
\label{fig:Dstst}
\end{center}
\end{minipage}
\hfill
\begin{minipage}[t]{0.35\textwidth}
\begin{center}
\epsfig{file=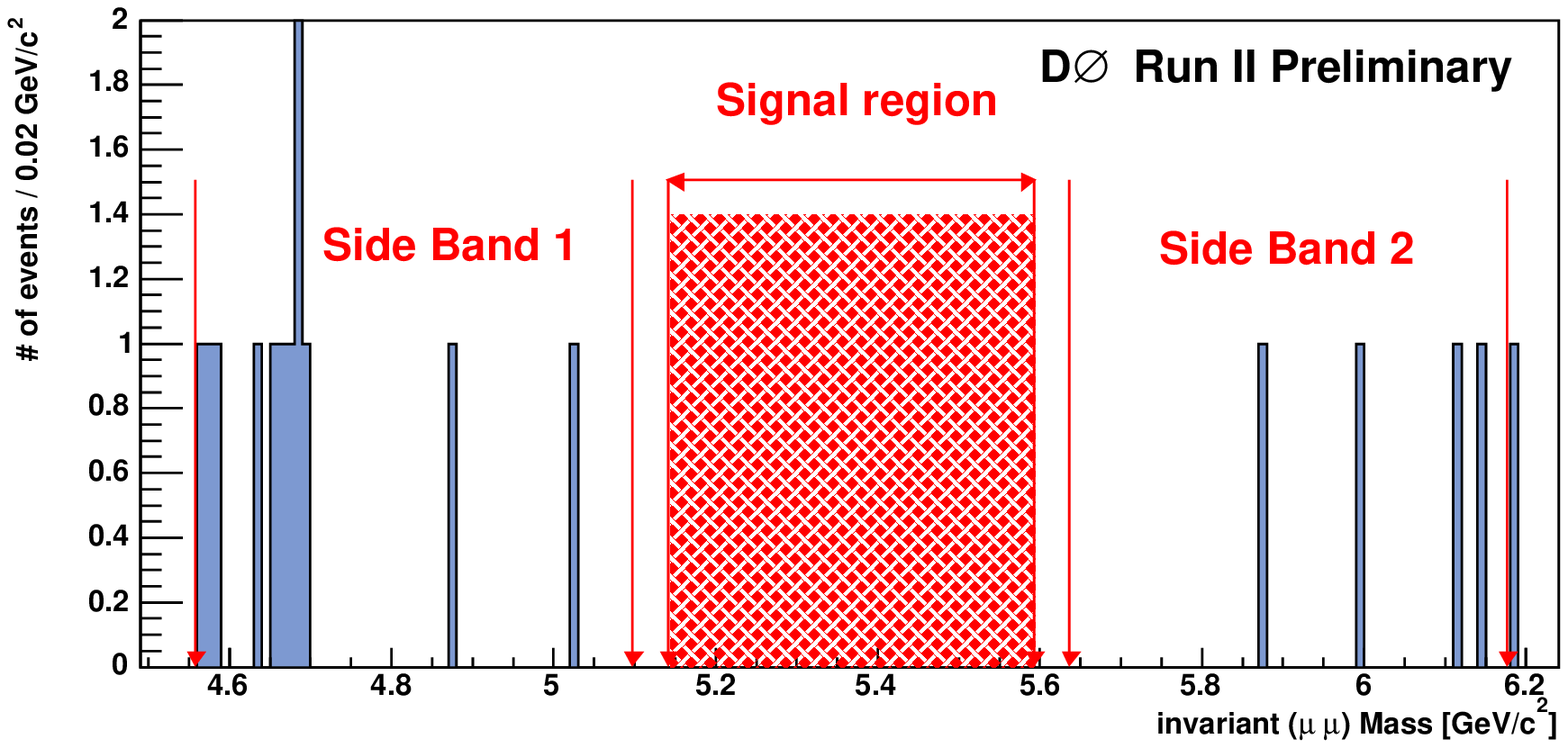,scale=0.35}
\caption{Definition of the signal and sideband regions in the $B_s \rightarrow \mu^+ \mu^-$ sensitivity analysis, and remaining background in 180~pb$^{-1}$ of data after preliminary cut optimisation.}
\label{fig:bsmumu}
\end{center}
\end{minipage}
\hfill
\end{figure}

As discussed, {\it e.g.}, in Refs.~\cite{bib:Jonas,bib:Buras,bib:Dedes} and references therein, the rare decay $B_s \rightarrow \mu^+ \mu^-$ is a promising window on possible physics beyond the Standard Model. We study the sensitivity of~D\O~to this decay using 180~pb$^{-1}$ of data. The main discriminating variables used to suppress background are the isolation of the muon pair, the transverse decay length~$|\vec{l}_{xy}|$ of the $\mu^+ \mu^-$~vertex w.r.t. the primary vertex, and the opening angle between~$\vec{l}$ and~$\vec{p}(\mu^+ \mu^-)$. The $\mu^+ \mu^-$~mass spectrum after cuts is shown in Fig.~\ref{fig:bsmumu}. Based on the number of events observed in the mass sidebands~(Fig.~\ref{fig:bsmumu}), we expect $7.3 \pm 1.8$~background events in the signal region~(Fig.~\ref{fig:bsmumu}), and, in the absence of an excess over the expected background, a limit of Br$(B_s \rightarrow \mu^+ \mu^-) < 1.0 \cdot 10^{-6}$ at $95 \; \%$~CL. This expected limit takes into account both statistical and systematic uncertainties. The expected signal has been normalised to an abundant sample of fully reconstructed $B^{\pm} \rightarrow J/\psi K^{\pm}$ decays. Possible contributions from $B_d \rightarrow \mu^+ \mu^-$ are expected to be small compared to $B_s \rightarrow \mu^+ \mu^-$, and are neglected here. The signal box has not been opened yet, {\it i.e.} the number of events in the signal region has not been looked at. The event selection is currently being reoptimised, and a very competitive limit on a slightly larger data sample is expected on a short timescale.

\section{Conclusion and outlook}

The large acceptance in terms of rapidity and transverse momentum of the charged particle tracking system and the muon system, including the muon trigger, make the upgraded Run~II D\O~detector an excellent tool for \B~physics. In this article we have presented selected physics results based on up to 250~pb$^{-1}$ of Run~II data. Large samples of semileptonic \B~decays have been accumulated using single muon triggers. The excellent impact parameter resolution of the tracking system allows precise measurements of the decay-time dependence of \B~decays, and one of the world's most precise measurements of $\tau(B^+)/\tau(B^0)$ has been performed. The muon reconstruction capabilities also allow pure flavour tagging, which is another key ingredient for the study of flavour oscillations and CP~violation. The good tracking efficiency down to very low transverse momenta has made possible the observation of semileptonic \B~decays to \Dstarstar~mesons. Large yields of hadronic decays of $b$~hadrons to $J/\psi \rightarrow \mu^+ \mu^-$ are obtained using di-muon triggers. Di-muon trigger data have also been used for a detailed study of the characteristics of the \X~state and a search for the rare decay~$B_s \rightarrow \mu^+ \mu^-$. Other hadronic decays are collected by triggering on single opposite-side muons. Events recorded in this way are particularly powerful for measurements that require flavour tagging, as the tag is provided by the muon that fired the trigger. In addition, the proper decay time distribution of these events is not biased by the trigger. Such events will be included in future analyses. Other improvements are expected from the future use of $dE/dx$~information from the silicon tracker, which provides kaon and proton identification at very low momenta. In combination with information from the tracking system, D\O's calorimeter is starting to be used to identify non-isolated low-$p_{\rm T}$ electrons. Together with the calorimeter, the preshowers (which provide finer granularity than the calorimeter) are used to identify $\pi^0$~decays. In summary, the exciting times of the competitive D\O~Run~II \B~physics program have started.

\end{document}